\documentclass[fleqn,usenatbib]{mnras}

\usepackage{newtxtext,newtxmath}

\usepackage[T1]{fontenc}

\DeclareRobustCommand{\VAN}[3]{#2}
\let\VANthebibliography\thebibliography
\def\thebibliography{\DeclareRobustCommand{\VAN}[3]{##3}\VANthebibliography}

\usepackage{graphicx}	
\usepackage{amsmath}	
\usepackage{xcolor}
\usepackage{float}
\usepackage{siunitx}
\usepackage{setspace}
\usepackage{geometry}
\usepackage{graphicx}
\usepackage{natbib}
\usepackage{pdfpages}
\usepackage{subcaption}
\usepackage[toc,page]{appendix}
\usepackage{threeparttable}

\newcommand{\Halpha}{H$\rm \alpha$ }

\title{Disc Tearing in a Be Star: Predicted 3D Observations}

\author[M.W. Suffak et al.]{
M.W. Suffak,$^{1}$\thanks{E-mail: msuffak@uwo.ca}
C.E. Jones,$^{1}$
A.C. Carciofi,$^{2}$
\\
$^{1}$Department of Physics and Astronomy, Western University, London, ON N6A 3K7, Canada\\
$^{2}$Instituto de Astronomia, Geof\'isica e Ci\'encias Atmosf\'ericas, Universidade de S\~ao Paulo, Brazil\\
}

\date{Accepted 2023 November 23. Received 2023 November 20; in original form 2023 September 20}

\pubyear{2023}

\begin{document}
\label{firstpage}
\pagerange{\pageref{firstpage}--\pageref{lastpage}}
\maketitle

\begin{abstract}
We build on our previous work involving smoothed particle hydrodynamic simulations of Be stars, by using the model that exhibited disc tearing as input into the three-dimensional Monte Carlo radiative transfer code {\sc hdust} to predict observables from a variety of viewing angles throughout the disc tearing process. We run one simulation at the start of each orbital period from 20 to 72 orbital periods, which covers two complete disc tearing events. The resulting trends in observables are found to be dependent on the relative position of the observer and the tearing disc. The \Halpha equivalent width, $V$ magnitude, and polarization can all increase or decrease in any combination depending on the viewpoint of the observer. The \Halpha line profile also displays changes in strength and peak separation throughout the tearing process. We show how the outer disc of the torn system can have a large effect on the \Halpha line profile, and also contributes to a wavelength-dependent polarization position angle, resulting in a similar sawtooth shape to the polarization percentage. Finally, we compare our predictions to Pleione (28 Tau) where evidence has suggested that a disc tearing event has occurred in the past. We find that our tearing disc model can broadly match the trends seen in Pleione's observables, as well as produce the two-component \Halpha lines observed in Pleione. This is the strongest evidence, thus far, of Pleione's disc having indeed experienced a tearing event.
\end{abstract}



\section{Introduction}
\label{sec:intro}

Classical Be stars are rapidly rotating B-type stars that, at some point, have had Balmer emission lines in their spectrum due to a gaseous disc that has emanated from the star itself \citep{Collins1987,Porter2003}. The exact process of this disc feeding is uncertain, however non-radial pulsations \citep{Baade2016} coupled with rapid rotation is the mechanism thought to be likely responsible \citep[for the most recent review, see][]{rivinius2013classical}. The evolution of classical Be star discs is currently best-described by the viscous decretion disc (VDD) model of \cite{Lee1991} where material is launched from the star via some unknown mechanism, and then is transported outwards via viscous effects.

In nearly all modelling efforts involving the VDD model, the Be star disc has been assumed to be in the equatorial plane of the central star. However, a number of recent works \citep{Cyr2017, Brown2019, Suffak2022} have used hydrodynamic models to investigate Be star discs that are tilted out of the equatorial plane due to the influence of a binary companion who's orbit is misaligned to the equatorial plane of the Be star. The most recent work of \cite{Suffak2023} examined the temperature structure and observables of Be star discs that are tilted away from the star's equatorial plane, and was the first to study tilted Be star discs using radiative transfer models. Such a scenario may be caused by supernova kicks, leading the resulting neutron star to have a misaligned orbit to the Be star spin axis \citep{Martin2009}. The misaligned binary star then provides a torque on the Be star disc, causing the disc to tilt out of the equatorial plane and undergo nodal precession about the binary angular momentum vector \citep{Martin2011}. Additionally, if the Be star is actively losing mass to the disc, then the disc feels a competing viscous torque trying to keep the disc aligned with the Be star equator. In special cases, the two competing torques can cause the disc to undergo tearing \citep{Nixon2013,Dogan2015,Suffak2022}.

\cite{Suffak2022} computed six simulations where the binary companion orbit was misaligned to the Be star equatorial plane, one of which exhibited the phenomenon of disc tearing. Over the course of 100 binary orbital periods, the disc underwent three separate disc tearing events, spaced roughly 30 orbital periods apart. After the disc tears, the inner and outer disc nodally precess independently, while the inner disc is still being injected with material from the Be star. This results in the inner disc growing radially and tilting back towards the Be star equatorial plane, while the outer disc undergoes large inclination changes during precession, and gradually dissipates. This process allows the disc tearing phenomenon to repeat multiple times.

The Be star Pleione (28 Tau, HD23862) has been a focus of recent publications \citep{Marr2022, Martin2022} due to evidence suggesting the disc of Pleione undergoes episodes of disc tearing, where the disc splits into separate components: an inner disc and an outer precessing disc. In their detailed modelling of Pleione, \cite{Marr2022} postulated that the disc tearing scenario of \cite{Suffak2022} is the best explanation for drops in the H$\alpha$ equivalent width (EW) and $V$ magnitude of Pleione. \cite{Martin2022} then followed this analytically, finding that a broken disc is the only way to match the precession period and disc size of Pleione. A two-disc scenario was also proposed earlier for Pleione by \cite{Tanaka2007}. For other studies of Pleione, see the introduction of \cite{Marr2022}.

In this work, we use the Smoothed Particle Hydrodynamics ({\sc sph}) simulation from \cite{Suffak2022} that showed disc tearing, as input into the 3-dimensional non-local thermodynamic equilibrium (3D nLTE) Monte Carlo radiative transfer code {\sc hdust} \citep{carciofi2006non}. This will produce fully 3D predictions of observables of the disc before, during, and after a disc tearing episode, and include a full velocity prescription from the {\sc sph} code, which is a first for Be star research. Not only do we compare this to the data from Pleione, but we show the general trends in observables that disc tearing may produce when viewed from a complete variety of orientations.

Section \ref{sec:sim_details} of this paper briefly summarizes the details of the disc tearing model of \cite{Suffak2022}, Section \ref{sec:tearing_obs} discusses the general trends seen in the disc tearing observables from a variety of vantage points, Section \ref{sec:Pleione_comp} shows the comparison between the disc tearing observables and Pleione's variability, and Section \ref{sec:discussion} provides our discussion and conclusions.

\section{Simulation Details}
\label{sec:sim_details}


The details about the {\sc sph} code used by \cite{Suffak2022}, as well as more specifics on the simulation setup, can be found in section 2.1 of their paper. The simulation initially starts with no disc, and at each subsequent timestep, equal-mass particles are injected into the Be star's equatorial plane with Keplerian velocities to grow the disc. Most of the injected mass falls back onto the Be star immediately, and a small fraction of the material stays to form the disc. The orbit of the binary companion is misaligned by $\ang{40}$ to this equatorial plane. Both the binary's orbital plane and the spin axis of the Be star are fixed throughout the simulation. For convenience, parameters used in the simulation from \cite{Suffak2022} are listed in Table \ref{tab:sph_sim_params}. We note that the orbital and stellar parameters are not the same as Pleione, as the goal of this paper is to produce observables of a broken disc system in general for broad applicability, but later we show similarities to the Pleione system.

\begin{table}
\centering
\caption{Simulation parameters for the {\sc sph} model from \citet{Suffak2022}.}
\label{tab:sph_sim_params}
\begin{threeparttable}[c]
\renewcommand{\TPTminimum}{\linewidth}
\makebox[\linewidth]{
\vspace{3mm}
\begin{tabular}[c]{cc}
\hline\hline
    Parameter & Value  \\
    \hline
    Be Star Mass & 8 $\rm M_\odot$ \\
     Be Star Radius & 5 $\rm R_\odot$ \\
    Be Star Effective Temperature & 20000 K \\
    Binary Star Mass & 8 $\rm M_\odot$ \\
    Binary Star Radius & 5 $\rm R_\odot$ \\
    Disc Temperature & 12000 K \tnote{a} \\
    Viscosity Parameter $\alpha$ & 0.1 \\ 
    Mass Injection Rate & $10^{-8} \, \rm M_\odot yr^{-1}$ \\
    Particle Injection Radius & $1.04 \, \rm R_p$ \\
    Binary Orbit Misalignment Angle & \ang{40} \\ 
    Binary Orbital Period & 30 days\\
    Binary Orbital Radius & 20.5 $\rm R_p$ \\
\hline
\end{tabular}}
\begin{tablenotes}
\item[a] The disc temperature is set to 60\% of the primary star's effective temperature. This value was found by \cite{carciofi2006non} to be the average temperature in the isothermal regions of the disc.
\end{tablenotes}
\end{threeparttable}
\end{table}

To adopt input from the {\sc sph} simulation for {\sc hdust}, we use an interface that takes the {\sc sph} particles at a desired timestep, and creates a computational grid in a format that {\sc hdust} accepts. We created the grid in spherical coordinates with 50 radial cells, 50 azimuthal cells, and 50 polar cells. We explored a larger number of cells for better resolution, but the differences were negligible. In a standard grid, cells are spaced logarithmically in the radial and polar direction, since a typical disc is concentrated nearest the equator and near the central star. However, because of the complexity of the disc configurations in this simulation, we have designed the grid such that the cells are more closely spaced when the disc density is highest in each vertical column. 

For the stellar parameters that {\sc hdust} requires, we input the source (primary star) as having a rotation critical fraction, $W$, of 0.7, which is average for Be stars \citep{rivinius2013classical}, a mass of 8 $\rm M_\odot$, polar radius of 4 $\rm R_\odot$ (consistent with $W\,=\,0.7$ to give an equatorial radius of 5 $\rm R_\odot$), and a luminosity of 2300 $\rm L_\odot$ (consistent with the primary's  temperature in Table \ref{tab:sph_sim_params}).

In \cite{Suffak2022}, this simulation was run for 100 orbital periods before the disc was allowed to dissipate. The simulation starts with a single disc that tilts progressively out of the equatorial plane as it becomes more massive. When tearing occurs, the inner disc tilts back to the equatorial plane, while the outer disc starts precessing. What follows is inner disc growth into a new full disc, while the outer, detached disc, partially dissipates and perhaps recombines with the inner disc. During these 100 periods, the disc tearing process occurred three times, with the first episode starting at 25 orbital periods, and then repeating approximately every 30 orbital periods.

\section{Analyzing the Tearing Disc}
\label{sec:tearing_obs}

\subsection{Disc Geometry}

\begin{figure}
    \centering
    \includegraphics[scale=0.35]{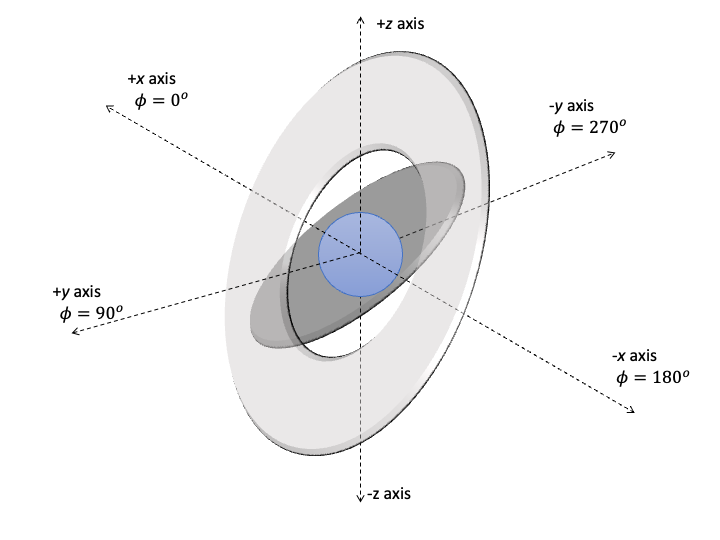}
    \caption{Schematic of the tearing disc system, showing the inner disc (dark grey) separated from the outer disc (light grey), both centered around the primary star (blue). In this frame, the $z$-axis is the spin axis of the Be star. The azimuthal angle ($\phi$, used for defining the obseving location) is also shown defined with $\phi=\ang{0}$ along the positive $x$-axis. Note that the relative sizes and inclination of the inner and outer discs to each other and to the central star would vary over time.}
    \label{fig:tearing_schematic}
\end{figure}

Before discussing observables of the tearing disc produced by {\sc hdust}, we first describe our coordinate orientation. A schematic of a torn disc is shown in Figure \ref{fig:tearing_schematic}, where we show an example of the inner and outer components of the disc when tearing occurs, although it is important to remember that this is just a schematic, and the inclinations of the two discs change over time. We define our coordinate system with two angles: the azimuthal angle $\phi$, which is in the range [\ang{0},\ang{360}) as shown in Figure \ref{fig:tearing_schematic}, and the polar angle $\theta$, defined from $\ang{0}$ at the pole of the star, to $\ang{90}$ at the equator of the star.

Overall, we compute observables at 152 different observing positions using ten $\theta$ values, every ten degrees from $\ang{0}$ (pole-on with the top of the star) to $\ang{180}$ (pole-on with the bottom of the star), and eight $\phi$ angles, every 45 degrees between $\ang{0}$ and $\ang{315}$. The observables are computed at the start of every period between 20 and 72 orbital periods, which covers two full disc tearing episodes, consisting of the disc tearing and then regrowing. During a disc tearing episode, the disc exists as two independent pieces: an inner disc whose inner boundary is still anchored at the Be star equator due to ongoing mass injection, and a detached outer ring, each inclined at different angles and nodally precessing independently. This means the projected area and projected velocity the observer sees from both disc pieces are very different at a given time, and can also change significantly over time. Visually, we determine the radius at which the disc tears is approximately 3 $R_*$. While mass injection is ongoing, the inner disc will feel competing torques from the binary as well as the addition of material, and the outer disc will only feel the binary torque and precess more quickly than the inner disc. Through analyzing the angular momentum vector of the inner and outer disc, we can calculate the longitude of the ascending node of each disc section throughout the simulation. The two discs do not undergo a full precession period as two separate pieces, so to find the precession periods of the inner and outer disc, we can look at the slope of the ascending node versus time when the two discs are completely separated (between 30 and 35 orbital periods), and compare what the period would be if the discs kept this rate of precession without change. Through this, we find the outer disc precesses approximately five times faster than the inner disc when the two discs are independently precessing.

It is not practical to go through the resulting observables for all 80 observing positions that we simulated, so we will focus on a few representative situations that best describe and explain the broad trends seen for all observing angles. In our disc tearing models, we chose to examine the H$\alpha$ line, $V$-band magnitude, $V$-band polarization percentage, and the corresponding polarization position angle (PA). 

\subsection{Photometry, Polarization, and Equivalent Width}
\label{sec:phot_pol_ew}

Throughout the simulation, as the disc cycles between tearing and regrowing, the observables also oscillate. The H$\alpha$ equivalent width (EW) can vary as much as 1 to 2 nm, the change in $V$ magnitude can be on order of hundredths or tenths of a magnitude, percent polarization amplitude can be as much as 1\%, and the polarization PA changes by tens of degrees. The exact amount of change in each observable is dependent on the orientation of the observer to the star and the tearing disc, as does the way in which the observables change with respect to each other. 

\begin{figure}
    \centering
    \includegraphics[scale=0.33]{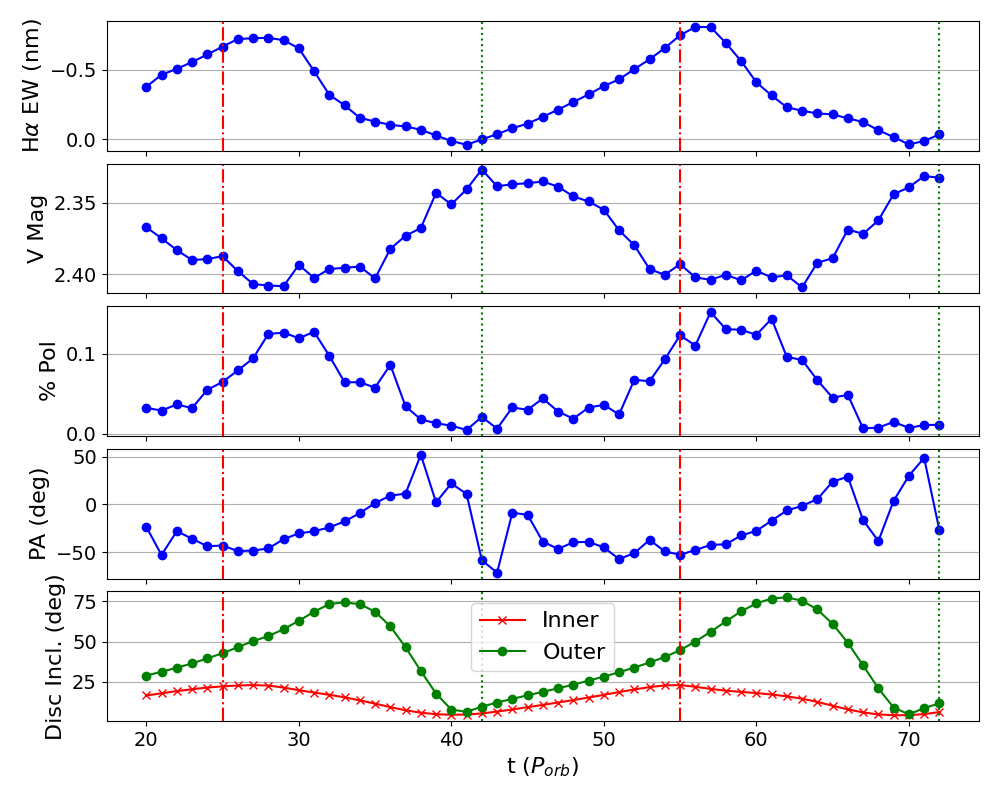}
    \caption{Top to bottom, plots of the computed \Halpha EW, $V$ magnitude, percent polarization, polarization PA, and average disc inclination relative to the observer of the inner ($< 3\, R_*$) and outer ($> 3\, R_*$) disc, for a pole-on observing angle of $\theta\,=\,\ang{0}$ and $\phi\,=\,\ang{90}$. The $x$-axis is in units of binary orbital periods. The vertical lines indicate times where the disc tearing begins (red, dot-dashed line) and where the disc is whole again (green, dotted line).}
    \label{fig:observables_10_90}
\end{figure}

Figure \ref{fig:observables_10_90} shows examples of this variability where we have plotted the \Halpha EW, photometry, polarization, and inclination of the inner ($< 3\, R_*$) and outer ($> 3\, R_*$) disc relative to the observer looking pole-on with the star $(\theta\,=\,\ang{0}$). Immediately before the disc tears it is tilted a small amount from the equatorial plane, accounting for the small, but non-zero, polarization signal. When the disc tears, the projected area of the disc that an observer would see from this pole-on angle decreases; as the optically thick \Halpha line emission strength is directly proportional to this area, the EW drops as disc tearing begins. After the disc tears, as the outer disc undergoes precession and is viewed more egde-on, the inner disc slowly realigns with the equatorial plane, which means pole-on observing angles will see the inner disc more face-on. As the $V$-band excess and polarization of a Be star disc originate from the inner portion of the disc \citep{Carciofi2011}, seeing the inner disc more face-on accounts for the brightening in $V$ magnitude, and the decrease in percent polarization as the polarization vectors would largely cancel due to symmetry. The variation in the PA depends on the direction of the tilt and orientation of the polarization axes in the simulation. While the inner disc is expected to cause much of the polarized flux, the outer disc also contributes to this flux, otherwise we would not see such large PA changes. In this case, the PA varies by about \ang{90} as the outer disc undergoes precession.

\begin{figure}
    \centering
    \includegraphics[scale=0.33]{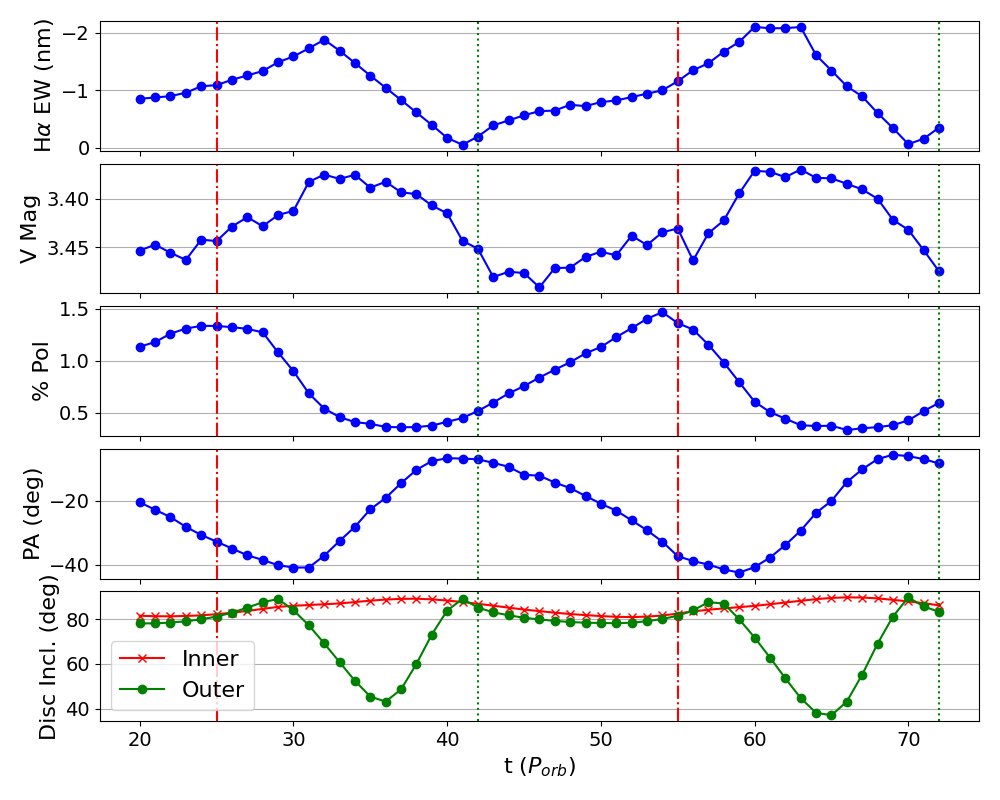}
    \caption{Same format as Figure \ref{fig:observables_10_90}, but for an observing angle of $\theta\,=\,\ang{90}$ and $\phi\,=\,\ang{45}$.}
    \label{fig:observables_90_45}
\end{figure}

As the observer moves to more equator-on views, tearing in the disc can lead to the opposite trends that are seen for pole-on angles. In many cases, the $V$ magnitude now decreases along with the H$\alpha$ equivalent width as the disc tears. The polarization percentage can both increase or decrease depending on how the inner disc orientation changes with respect to the observer. In some other cases, the EW or $V$ magnitude can go through multiple increases and decreases between the disc tearing and regrowing. An example of the trends when tearing is viewed equator-on is shown in Figure \ref{fig:observables_90_45}.

\subsubsection{Spectropolarimetry}

\begin{figure}
    \centering
    \includegraphics[scale=0.35]{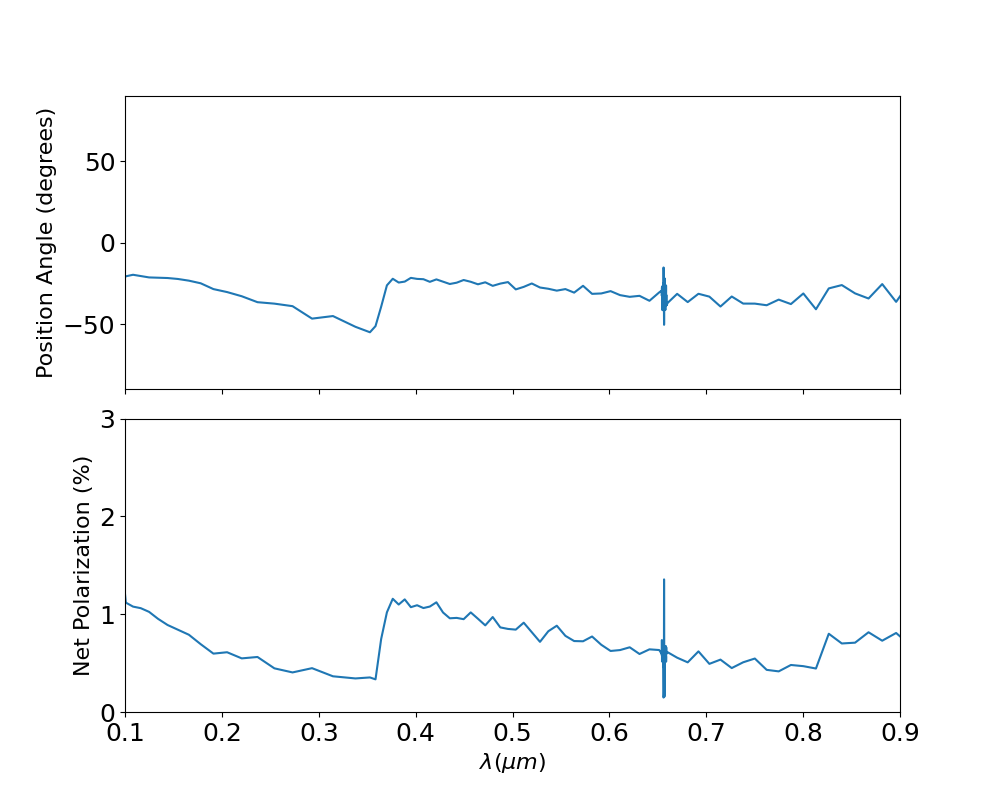}
    \caption{Polarization PA (top), and net polarization percentage (bottom), versus wavelength, for the tearing disc at 31 $\rm P_{orb}$ (shortly after disc tearing). Viewed from $\theta\,=\,\ang{90}$, $\phi\,=\,\ang{90}$.}
    \label{fig:mod168_PA}
\end{figure}

\begin{figure*}
    \centering
    \includegraphics[scale=0.35]{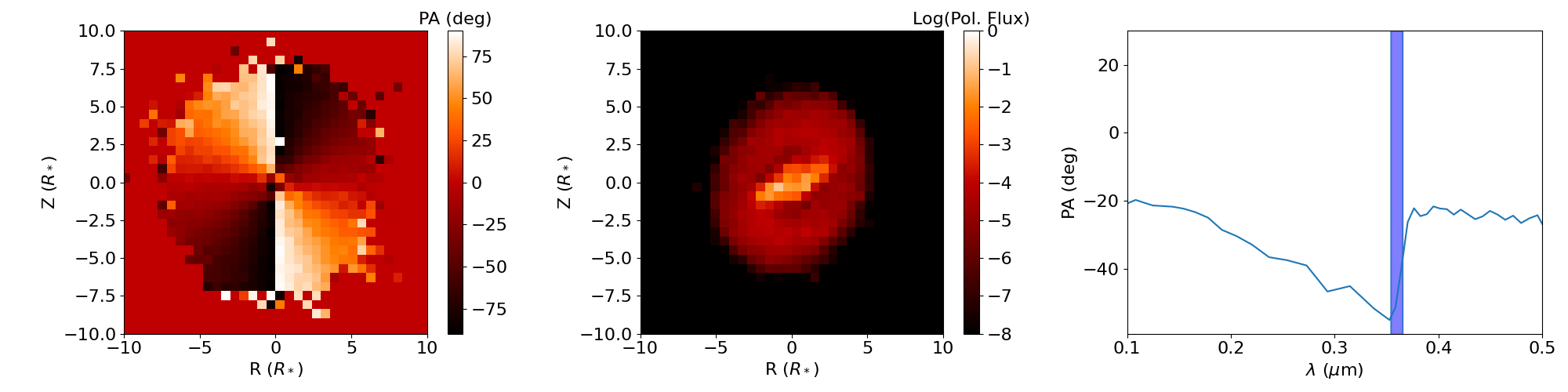}
    \includegraphics[scale=0.35]{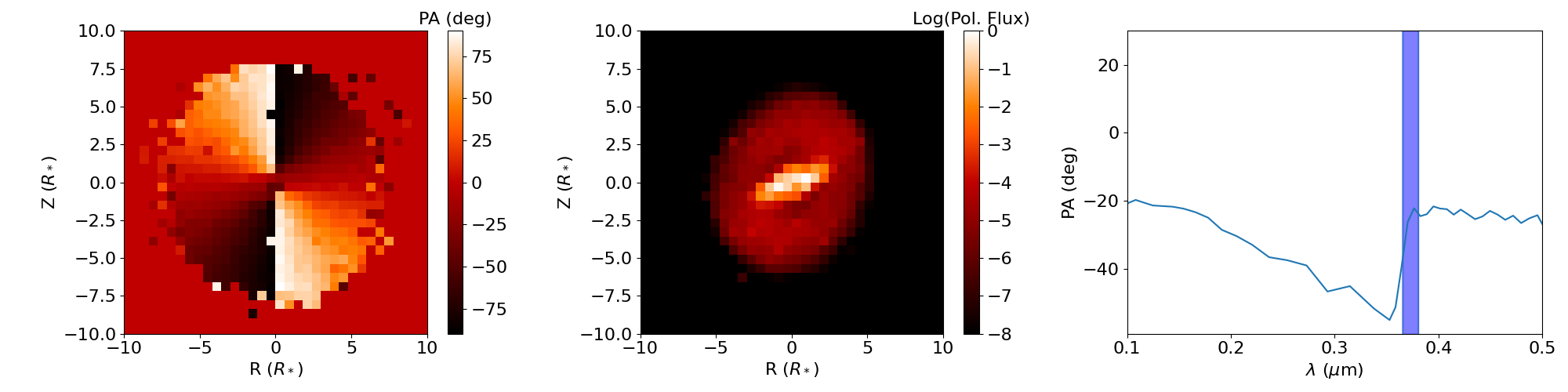}
    \caption{Left to right, the disc image coloured by pixel polarization angle, disc polarized flux, and polarization PA versus wavelength, with the blue highlighted region corresponding to the wavelength the two other panels are sampled from. The top plot is just before the Balmer jump, and the bottom plot is just after the Balmer jump. The viewing angle is $\theta\,=\,\ang{90}$, $\phi\,=\,\ang{90}$.}
    \label{fig:mod168_PA_explanation}
\end{figure*}

The polarization values presented in Figures \ref{fig:observables_10_90} and \ref{fig:observables_90_45} are averaged over the $V$ band, so the wavelength-dependent nature of Be star polarization is lost when computing these values. It is well known that the amount of polarization from a Be star disc takes on a wavelength dependent sawtooth-shaped pattern, due to the imprint of varying hydrogen opacity on the polarized light \citep[see][for example]{Haubois2014}, with jumps in the polarization occurring at the various hydrogen series boundaries. In examining spectropolarimetry from the tearing disc model, we find this behaviour in the polarization percentage, however, for some observing angles, we also find a sawtooth pattern in the polarization PA right after disc tearing, when the detached disc maintains its maximum mass (Figure \ref{fig:mod168_PA}). This behaviour is unexpected and would never be seen in a normal flat Be star disc since all of the disc material would be in the same plane, and the large majority of the polarization comes from the very innermost disc. However, it is not a surprising feature when considering both inner and outer discs contribute to the polarization.

In Figure \ref{fig:mod168_PA_explanation}, the left panel shows the disc image produced by {\sc hdust}, with each pixel coloured according to its PA, calculated by
\begin{equation}
    {\rm PA} = \frac{1}{2}\arctan\frac{U}{Q}.
\end{equation}
The middle panel shows this same disc image, but with the pixels coloured by polarized flux, and the rightmost panel highlights the wavelength range used in the left and middle images. The middle panel is essentially a map of the disc density. In comparing the left and middle panel, we can see that the outer disc has a much different polarization PA than the inner disc due to their opposing orientations. The middle panel also allows us to see that right before the Balmer jump (top row in Figure \ref{fig:mod168_PA_explanation}), when the PA is at a negative maximum, there is much less scattered light from the inner disc than right after the Balmer jump (bottom row in Figure \ref{fig:mod168_PA_explanation}), while the amount of scattered light from the outer disc remains the same. This means the outer disc is contributing a larger proportion of the total scattered light before the Balmer jump, enough to significantly change the PA from its value when the inner disc dominates. Incidentally, Figure \ref{fig:mod168_PA_explanation} demonstrates the statement made above that the inner disc contributes more to the polarized flux than the outer disc, but also shows that the contribution of the latter is not negligible.

\subsection{The H$\alpha$ Line}
\label{sec:sim_Halpha_line}

\begin{figure*}
    \centering
    \includegraphics[scale=0.35]{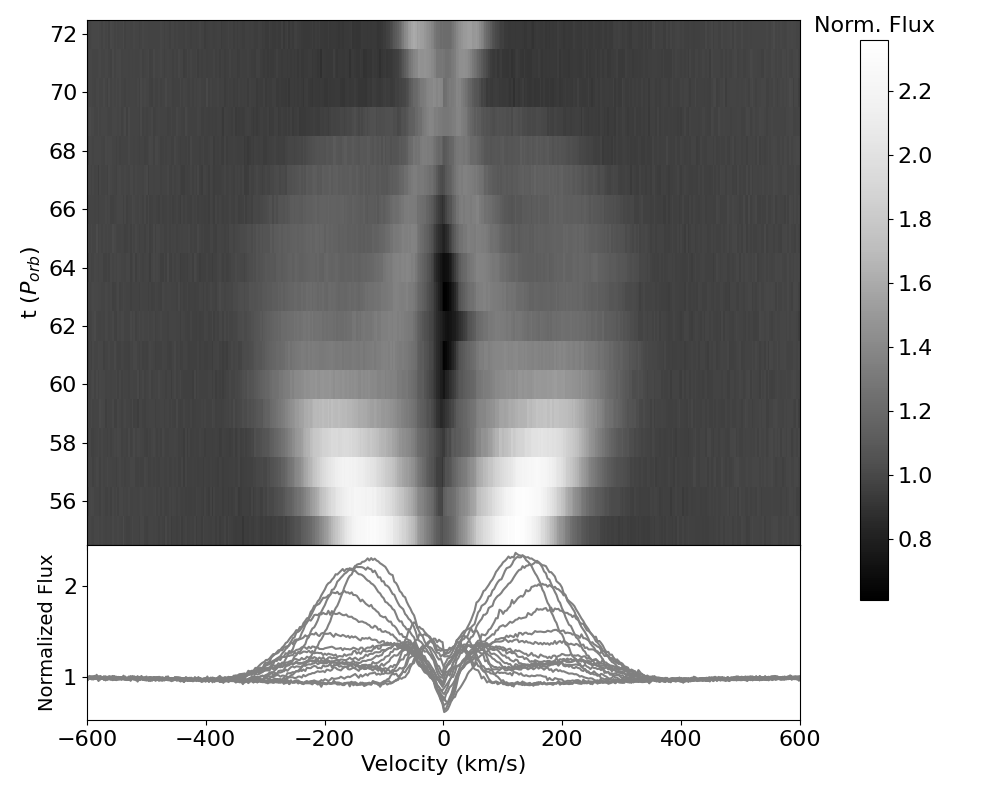}
    \includegraphics[scale=0.35]{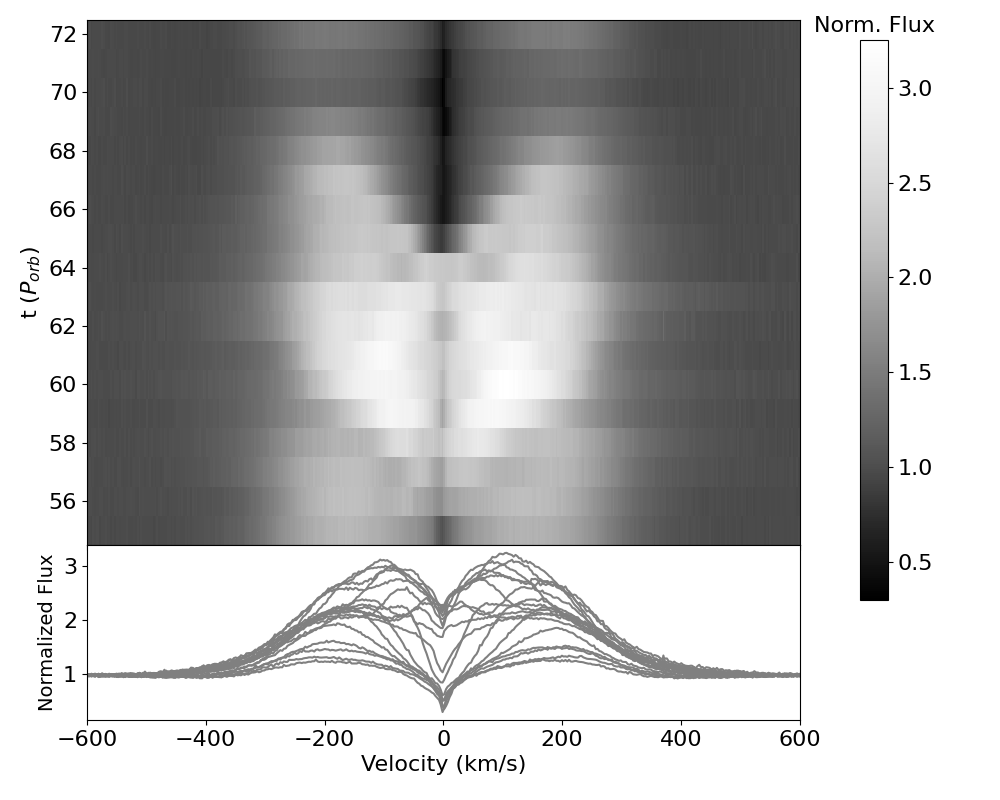}
    \caption{(Top) Dynamical \Halpha spectra of the tearing disc model from 55 to 72 orbital periods.(Bottom) \Halpha line profiles used to make the top plot. The colourbar indicates the normalized flux at each velocity along the x-axis. A velocity of zero indicates \Halpha line center. (Left) The observing angle is $\theta\,=\,\ang{0}$ and $\phi\,=\,\ang{90}$. (Right) The observing angle is $\theta\,=\,\ang{90}$ and $\phi\,=\,\ang{45}$.}
    \label{fig:dyn_spectra_10_90}
\end{figure*}

We also examined the H$\alpha$ line profile, with respect to how the disc tearing process can change its shape. Overall the H$\alpha$ line shape keeps a largely ``normal" double-peaked profile, however the line strength and peak separation can change as the disc tilts and precesses before/during/after the tearing process, depending on the orientation relative to the observer. Figure \ref{fig:dyn_spectra_10_90} shows examples of this for a pole-on and equator-on viewing angle, by plotting the dynamical H$\alpha$ spectrum across an entire episode of disc tearing and recombination. This Figure shows how not only the line strength, but also the peak separation and strength/width of the central reversal, can vary throughout the disc tearing process as different portions of the Doppler shifted disc material can be viewed from the observer's vantage point. 

\begin{figure}
    \centering
    \includegraphics[scale=0.35]{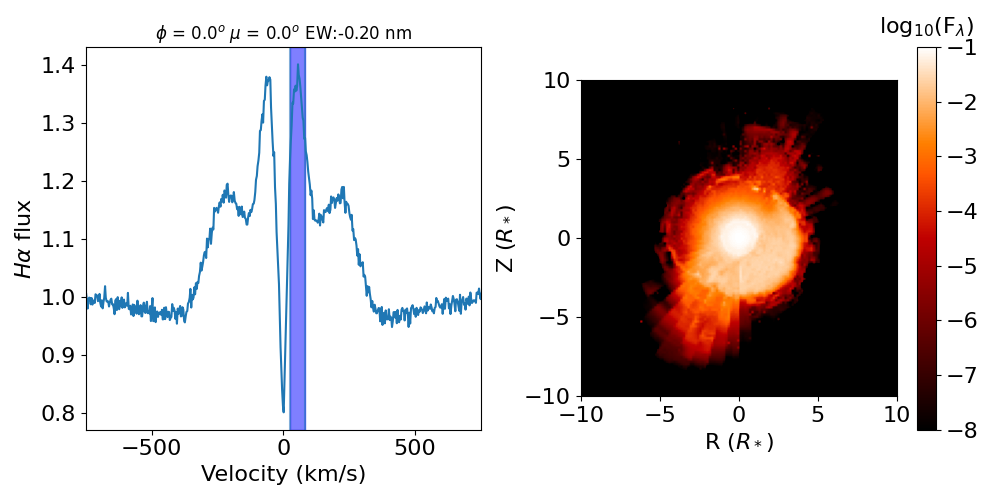}
    \includegraphics[scale=0.35]{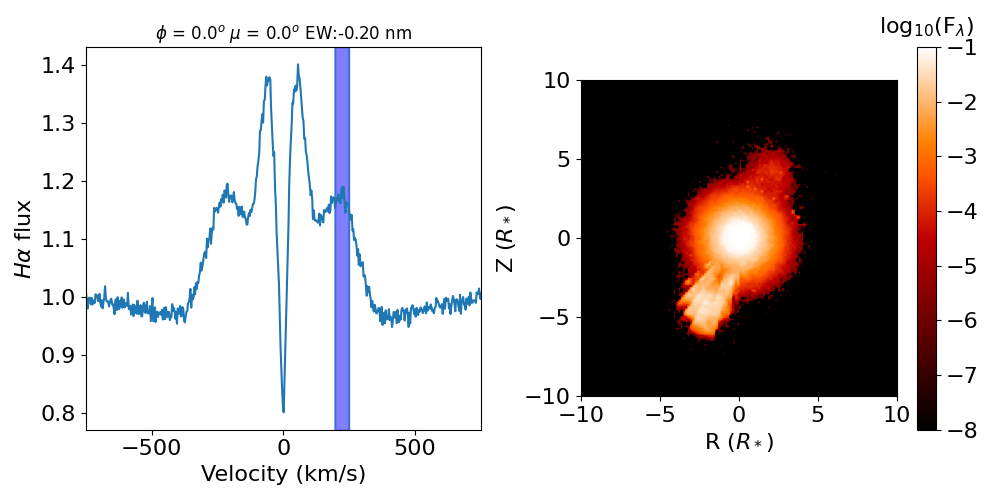}
    \caption{A two-component \Halpha line (left) along with the corresponding {\sc hdust} image (right), showing the emission in the velocity bin highlighted in blue on the line profile. The top plot is for the red peak of the line, while the bottom plot is for the red wing of the line as indicated on the left panels.}
    \label{fig:Halpha_vel_images}
\end{figure}

For particular observing orientations, at certain times, the H$\alpha$ line presents a ``two-component" profile, where the line is clearly composed of a broad and narrow component, that wouldn't be seen in an ordinary disc. Figure \ref{fig:Halpha_vel_images} shows an example of one of these two-component lines on the left, with the corresponding disc image produced by {\sc hdust} in the H$\alpha$ velocity bin that is highlighted in blue on the line profile in the left panel of the Figure. From this, we can see that the inner double peaks in the line originate from the inner disc, while fast moving wings of the two-component line are due to the outer ring that has separated from the inner disc during the tearing process. This is consistent with the geometry of the disc from this observing angle, as the inner disc is close to face-on and thus has lower projected velocities than the outer ring which is edge-on with the observer. So we can say that the line profile is in fact a combination of two profiles, one produced by the inner disc, and one from the outer disc. The visible separation between the broad and narrow components in the line profile are strongest just after the disc tears, and tend to vanish as the disc regrows and the outer disc partially dissipates and perhaps recombines with the rest of the disc.

\subsection{Role of the Outer Disc}
\label{sec:outer_v_inner_disc}

The conclusion in the previous paragraph naturally raises the question: how much does the outer disc contribute to the \Halpha emission line, as well as the rest of the resulting spectrum? To test this, we compute an {\sc hdust} simulation at 65 orbital periods - when the disc is completely torn into two parts - and remove the outer disc from the computational grid, meaning this simulation will produce observables from only the star and inner disc. 

\begin{figure}
    \centering
    \includegraphics[scale=0.35]{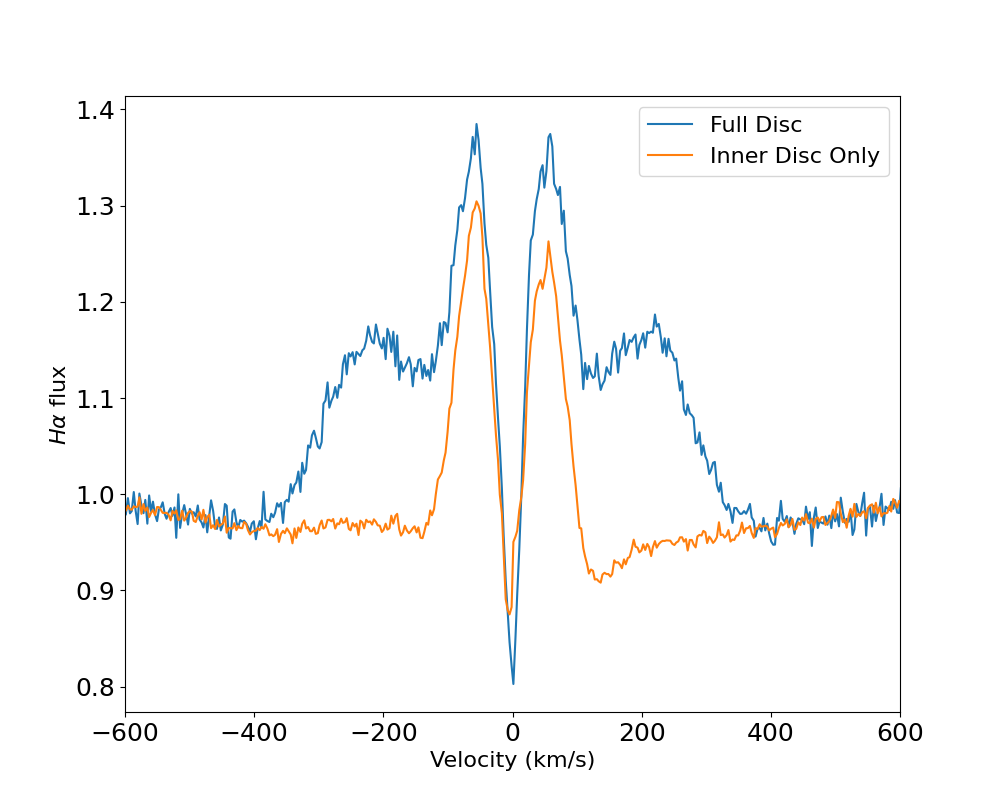}
    \caption{\Halpha line computed with {\sc hdust}, viewed directly pole-on with the star and at 65 orbital periods, of the entire system (blue) and with the outer disc removed (orange).}
    \label{fig:Ha_outer_removed}
\end{figure}

\begin{figure}
    \centering
    \includegraphics[scale=0.3]{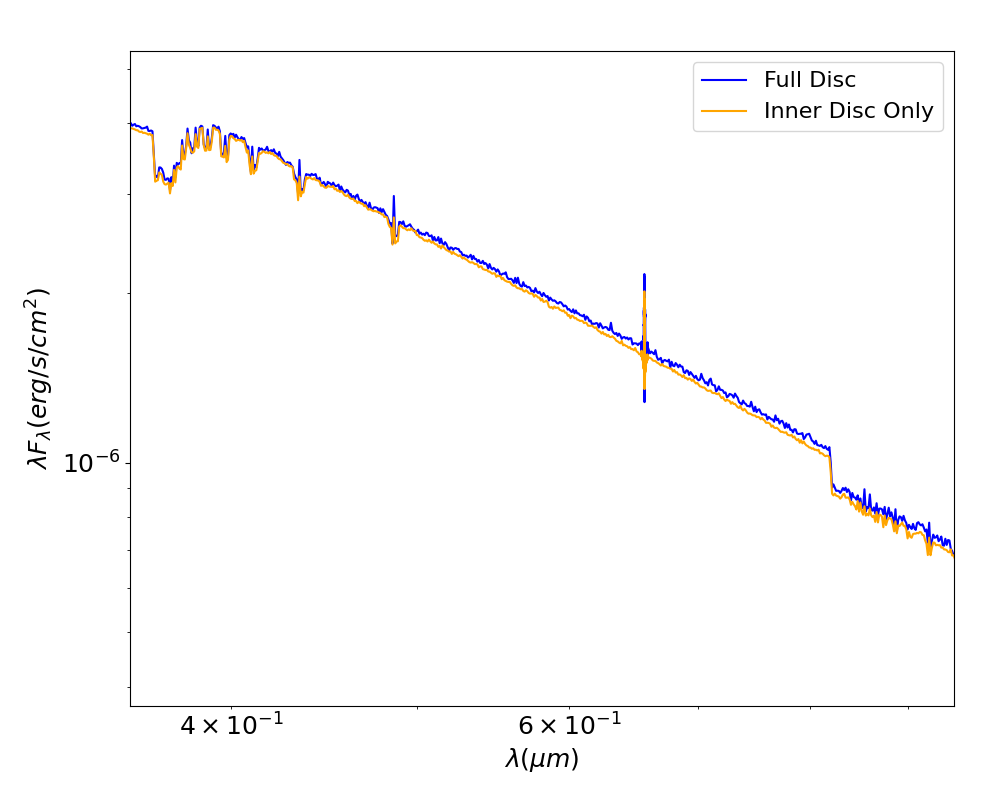}
    \caption{Visible and near-infrared spectrum computed with {\sc hdust}, viewed directly pole-on with the star and at 65 orbital periods, of the entire system (blue) and with the outer disc removed (light blue).}
    \label{fig:spectrum_outer_removed}
\end{figure}

The results of this test are shown in Figures \ref{fig:Ha_outer_removed} and \ref{fig:spectrum_outer_removed}, where we show, for a pole-on view, a comparison of the \Halpha line, as well as the visible and near-infrared spectrum, between the simulation with the entire torn disc system, and simulation containing just the inner disc and central star. 

When analyzing the results of the \Halpha line, one must remember that the shape of the line is largely due to the relative velocities between the disc and the observer, which will vary with changing inclination of the disc to the observer. In Figure \ref{fig:Ha_outer_removed}, we can see that removing the outer disc has a large effect on the emission line, removing the wings of the line entirely, while the inner double-peaks of the line remain approximately unchanged. This reaffirms our previous findings that, from this observing angle, the outer disc was responsible for the higher velocity wings of the line, while the inner disc produces the inner double-peaked line structure. It is important to understand that for other observing angles, the reverse may be true and the outer disc may be responsible for the lower velocity portions of the line, and the inner disc responsible for the wings, depending on the vantage point. 


In Figure \ref{fig:spectrum_outer_removed}, we see that removing the outer disc has very little effect on the visible and near-infrared spectrum. The same is seen in the polarization signature as well, with the exception that the PA no longer has the sawtooth shaped wavelength dependence. This is expected as the visible and near-infrared excess of Be star discs, as well as the polarization, is produced within the first few stellar radii of the disc, nearest the star and where the disc is densest \citep{rivinius2013classical, Haubois2012, Haubois2014}. 

\section{Comparison of the Disc Tearing Model to Pleione}
\label{sec:Pleione_comp}

The data on Pleione collected by \cite{Marr2022} spans a timeframe of decades and includes $V$-band photometry, \Halpha EW, percent polarization, as well as polarization PA. The authors also published \Halpha line profiles from 2005 to 2019. To facilitate comparison between our simulation results and the data of Pleione, we obtained the \Halpha spectra from the online journal version of \cite{Marr2022}, and secured the data presented in their Figure 14 (private communication).

Pleione shows a sharp drop in \Halpha EW and $V$ magnitude, while the polarization percentage is beginning to increase, around 2007 (MJD $\approx$ 54000). This is the behaviour that \cite{Marr2022} was unable to match with conventional disc models, and then hypothesized that disc tearing was occurring.

\begin{figure}
    \centering
    \includegraphics[scale = 0.35]{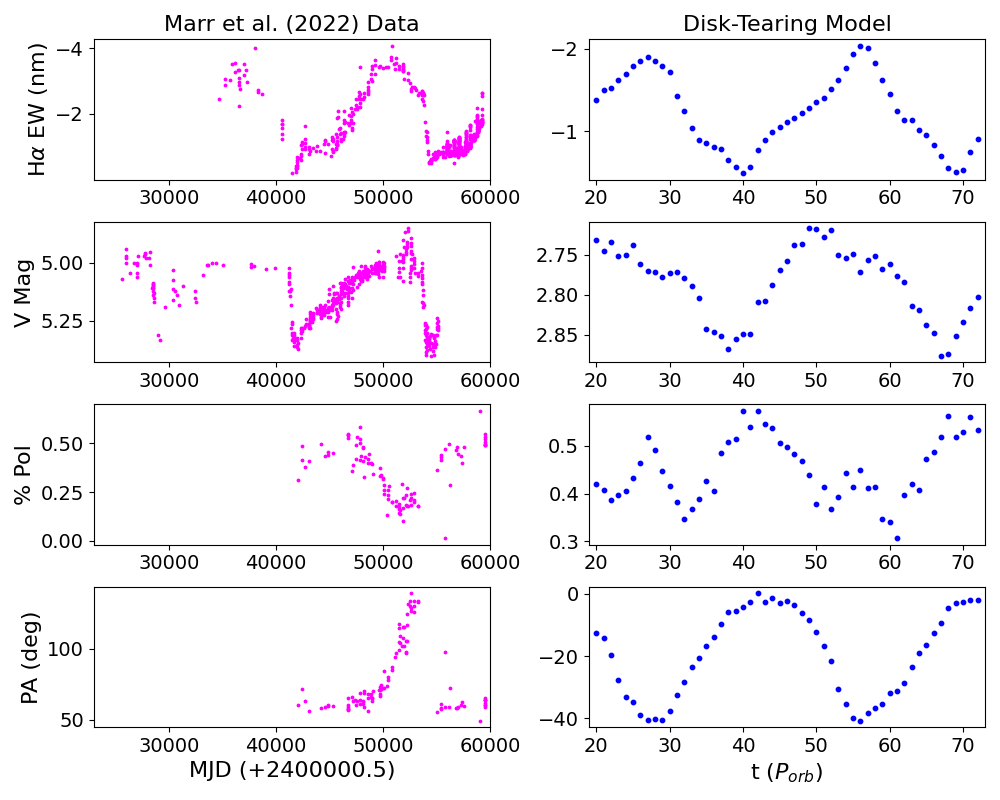}
    \caption{\Halpha EW, $V$ magnitude, percent polarization, and polarization PA data of Pleione from \citet{Marr2022} (left, top to bottom), and predictions from our disc tearing model from 20 to 72 orbital periods viewed from $\theta\,=\,\ang{60}$ and $\phi\,=\,\ang{0}$.}
    \label{fig:pleione_data_comp}
\end{figure}

As stated in Section \ref{sec:tearing_obs}, for some observing angles close to equator-on with the star, we see these exact trends of EW and $V$ magnitude decreasing while polarization increases. Figure \ref{fig:pleione_data_comp} shows a side-by-side comparison of Pleione's data from \cite{Marr2022} with the observables of our disc tearing model computed at $\theta\,=\,\ang{60}$ and $\phi\,=\,\ang{0}$. While the drops in \Halpha EW and $V$ magnitude from the disc tearing model are not as steep as those in Pleione, the overall trends are the same, and in the case of EW, the trend is remarkably similar before this sharp drop as well. 

\begin{figure}
    \centering
    \includegraphics[scale = 0.35]{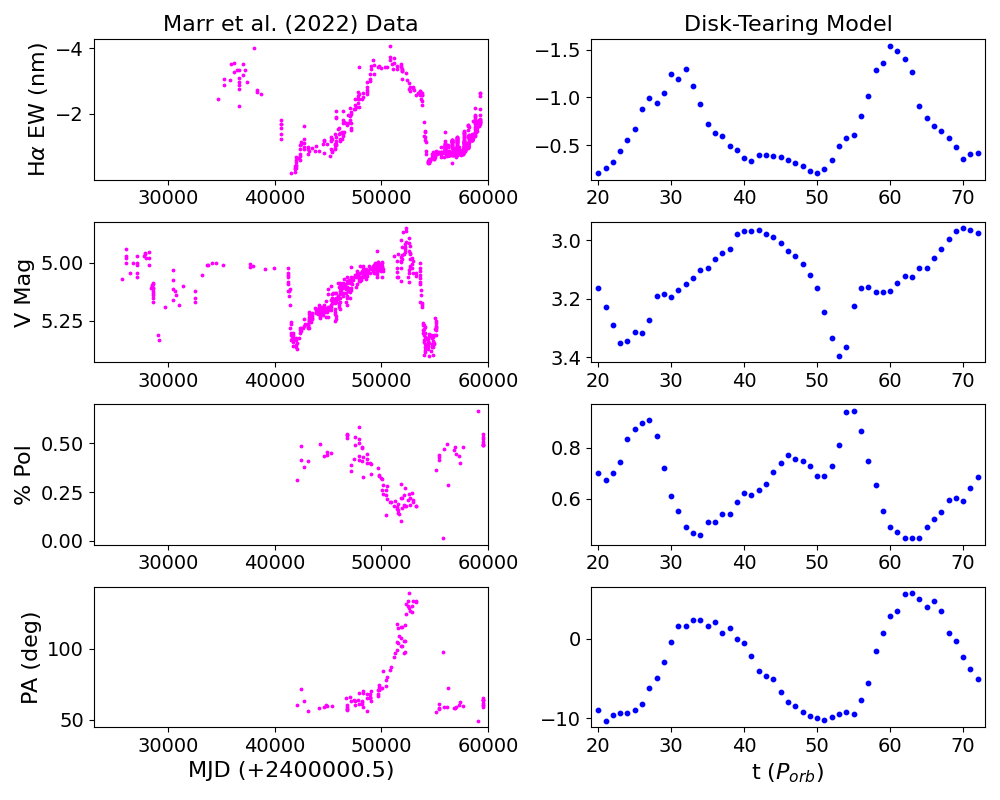}
    \caption{Same as Figure \ref{fig:pleione_data_comp}, but the model is viewed from $\theta\,=\,\ang{60}$ and $\phi\,=\,\ang{135}$. }
    \label{fig:pleione_data_comp_zoom}
\end{figure}

Other observing angles in our simulation are able to match some features in Pleione as well. We show a second example in Figure \ref{fig:pleione_data_comp_zoom}, viewed from the same polar angle as Figure \ref{fig:pleione_data_comp}, but from $\phi\,=\,\ang{135}$. While the decrease in $V$ magnitude and \Halpha EW do not occur simultanesouly here, the system shows a much steeper drop in photometry and polarization. The trends in EW and PA do not differ significantly from Figure \ref{fig:pleione_data_comp} and still match those seen in Pleione. The viewing angle of $\theta\,=\,\ang{60}$ corresponds to the same stellar inclination that \cite{Marr2022} identified for Pleione. It is worth noting that these trends arealso observed for polar angles greater than $\ang{60}$, spanning multiple azimuthal angles.


\begin{figure}
    \centering
    \includegraphics[scale = 0.35]{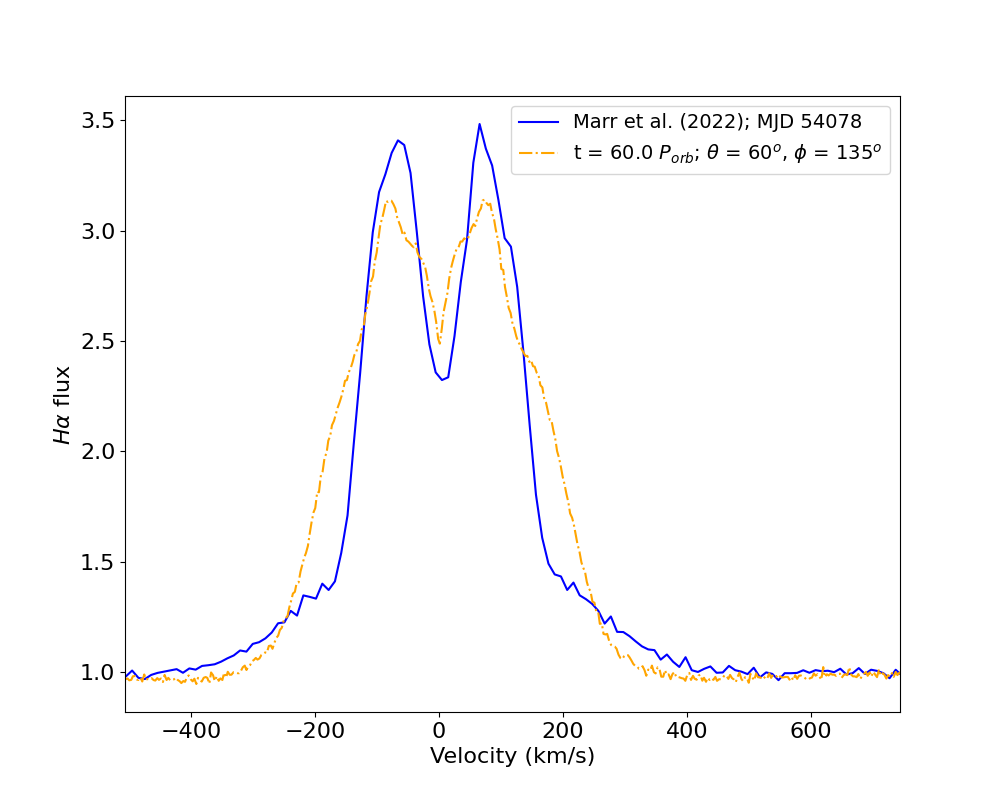}
    \caption{Comparison between one of Pleione's \Halpha lines from \citet{Marr2022} that shows the two-component line profile, and one such line from our disc tearing model, taken at 60 $\rm P_{orb}$, viewed from $\theta\,=\,\ang{60}$ and $\phi\,=\,\ang{135}$.}
    \label{fig:Ha_line_comp}
\end{figure}

From 2005 to 2019, the \Halpha line in Pleione transitioned from strongly double-peaked in 2005 to a shell-line post-2007. In the time between these two profiles, the line displays the same two-component behaviour that we discovered in our simulations, as discussed in Section \ref{sec:sim_Halpha_line}. Figure \ref{fig:Ha_line_comp} shows the \Halpha line in our simulation seen from $\theta\,=\,\ang{60}$, $\phi\,=\,\ang{135}$, at a time when the disc is beginning to tear, overlayed with Pleione's \Halpha line from 2006 from \cite{Marr2022}. We see that the shape of the lines are qualitatively the same, with slight differences in the peak heights and width of the line. In our simulation, the lines are expected to be wider than those of Pleione due to a more massive star than Pleione adopted in this work. This means the material around the star is rotating much faster in our simulation than it would around Pleione.

\section{Discussion and Conclusions}
\label{sec:discussion}

In this work, we used the disc tearing {\sc sph} simulation from \cite{Suffak2022} as input into the radiative transfer code {\sc hdust} to examine the trends in observables that a tearing disc would produce. This is the first time a true 3D {\sc sph} and radiative transfer model has been used to examine observations while a disc tears. We produced simulated observables for the disc tearing model at the start of every orbital period, from 20 to 72 orbital periods and calculated the $V$-band magnitude, percent polarization, polarization PA, and the \Halpha EW, in addition to analyzing the \Halpha line shape.

Much like the findings of \cite{Suffak2023}, who examined the temperature and observables of flat (non-warped) tilted discs, the variation in the observables of a tearing disc are entirely dependent on the orientation of the disc with respect to the observer. When the disc tears, any of the observables we examined could increase or decrease with some correlated or anti-correlated with other observables (see Figures \ref{fig:observables_10_90} and \ref{fig:observables_90_45}). However, we do find that usually the start of the disc tearing process corresponds to a maximum or minimum in all observables we examined. Furthermore, for a majority of the observing angles, the \Halpha EW seems to peak near the start of disc tearing, and then decreases. This is due to most observing angles seeing a smaller projected area of the disc when the disc tears. With respect to the \Halpha line, we find throughout the disc tearing process the line strength and peak separation can change significantly. This is expected, as tearing in the disc changes the projected area and relative velocities of the disc no matter what angle the disc is being viewed from.

We also find that the \Halpha line can present a ``two-component" line profile, where there is a clear separation between the broad line wings and the narrow inner double-peaked structure. We find that the outer disc can significantly contributes to the wings of the line in these cases (see Figure \ref{fig:Ha_outer_removed}), which shows how mapping each line component to a particular disc portion is difficult without a good grasp of the ensuing disc geometry. The outer disc is can also affect the polarization PA near the hydrogen series limits, causing the PA to have a sawtooth shape wavelength dependence, however it minimally contributes to the continuum flux and polarization magnitude. This is consistent with past publications of Be stars showing that the inner most area of a Be star disc is responsible for continuum and polarized emission, since this is usually where there is a large amount of UV radiation and electrons for scattering due to this region having the largest density.

The outer disc also does not contribute equally to all emission lines produced by the disc. It is well known that emission lines of Be stars have different emitting regions \citep[see figure 3 of][for example]{Wisniewski2007}, with the \Halpha line having one of the largest emitting volumes \citep{rivinius2013classical}. Because many emission lines form close to the star (especially IR lines), disc tearing is not expected to effect all lines equally, and many will not show the same ``two component" profile seen in \Halpha. This could be useful information, allowing the radius at which the disc is tearing to be predicted. The geometry of disc warping and tearing in combination with different emitting volumes may also cause some emission lines to be in a shell phase while others may be in a normal Be-phase. This is certainly something that deserves a thorough investigation, but is beyond the scope of this study.

We then examined the similarities between our predictions and observations of the Be star Pleione (28 Tau). \cite{Marr2022} proposed that the long-term behaviour in Pleione could be due to disc tearing, but at the time did not have the 3D tools developed to investigate this hypothesis. In comparing our model to Pleione, we find that the observables produced by our disc tearing model viewed from $\theta\,=\,\ang{60}$, $\phi\,=\,\ang{0}$, can roughly match the trends that have been observed in Pleione. From Figure \ref{fig:pleione_data_comp}, we can see that the \Halpha EW and $V$ magnitude drop at approximately the same time in our models, as they do in Pleione, while the percent polarization reaches a minimum and begins to increase, and the PA has peaked, beginning to decrease. The drops that we see from this perspective in the EW and photometry are not seemingly as sharp as those observed in Pleione, however the observing angle $\theta\,=\,\ang{60}$ and $\phi\,=\,\ang{135}$, seen in Figure \ref{fig:pleione_data_comp_zoom}, can produce these steeper drops in $V$ magnitude and polarization, despite the photometry and EW not decreasing at the same time. In examining the \Halpha line, we also find a similar two-component emission line that Pleione has (Figure \ref{fig:Ha_line_comp}), at  $\theta\,=\,\ang{60}$ and $\phi\,=\,\ang{135}$. This inclination is significant as \cite{Marr2022} determined the stellar inclination of Pleione to be $\ang{60}$ through modelling Pleione when it did not have a disc.

We caution that the {\sc sph} simulation parameters and stellar parameters are not the same as the Pleione system; however, the general qualitative agreement does add additional support to the hypothesis that Pleione undergoes dis tearing events. The \Halpha line and other observables will change with different model parameters such as the mass ratio, mass-loss rate, the viscosity parameter $\alpha$, and the binary orbit inclination, all of which also change the tearing radius in the disk \citep[see equation 8 of][]{Dogan2015}. So there are likely many combinations of parameters that could produce similar observable trends and line shape as Pleione at a consistent inclination. We also simulated systems with the same parameters but a lower mass ratio than the one presented here, and find disc tearing does occur for a mass ratio of a half, with even lower mass ratios still to be investigated. 

The results of this paper present the first study at the observables of a tearing Be star disc, using a 3D {\sc sph} and radiative transfer model in combination. The trends presented here will be invaluable in providing a baseline for evidence of disc tearing in binary systems with discs. Our example of this using the Be star Pleione provides the most conclusive evidence to date that disc tearing is occurring in Pleione, despite different simulation parameters used in this work. Our study opens the door to understand a broad range of astrophysical objects with discs for the first time, allowing complex disk systems to be investigated with fully 3D modelling techniques, and is applicable to other investigations where circumstellar discs tilt, become unstable, and tear.

\section*{Acknowledgements}

The authors thank the anonymous referee for their comments and suggestions which helped improve the paper. C.E.J. acknowledges support through the National Science and Engineering Research Council of Canada. A. C. C. acknowledges support from CNPq (grant 311446/2019-1) and FAPESP (grants 2018/04055-8 and 2019/13354-1). This work was made possible through the use of the Shared Hierarchical Academic Research Computing Network (SHARCNET).

\section*{Data Availability}

No new data was generated in this work.

\bibliographystyle{mnras}
\bibliography{aastexBeStarbib}

\bsp	
\label{lastpage}
\end{document}